\documentclass[
prd
,showpacs,amssymb,superscriptaddress,aps
]{revtex4}
\usepackage{graphicx}
\usepackage{color}
\input{epsf}

\usepackage{amsmath,amssymb}
\usepackage{bm}
\usepackage{times}
\usepackage{ulem}

\newcommand{\dalm}{\kern1pt\vbox{\hrule height 0.9pt\hbox{\vrule width 0.9pt
\hskip 2.5pt\vbox{\vskip 5.5pt}\hskip 3pt\vrule width 0.3pt}\hrule height 0.3pt}
\kern1pt}

\begin{document}



\title{Gravitational wave asteroseismology on cooling neutron stars}

\author{Hajime Sotani}
\email{sotani@yukawa.kyoto-u.ac.jp}
\affiliation{Astrophysical Big Bang Laboratory, RIKEN, Saitama 351-0198, Japan}
\affiliation{Interdisciplinary Theoretical \& Mathematical Science Program (iTHEMS), RIKEN, Saitama 351-0198, Japan}

\author{Akira Dohi}
\affiliation{Department of Physics, Kyushu University, Fukuoka, 819-035, Japan}
\affiliation{Interdisciplinary Theoretical \& Mathematical Science Program (iTHEMS), RIKEN, Saitama 351-0198, Japan}

\date{\today}

\begin{abstract}
We examine the gravitational wave frequencies from neutron stars during thermal evolution, adopting the relativistic Cowling approximation. We particularly focus on the neutron star models, in which the direct Urca (rapid cooling process) does not work, without the superfluidity and superconductivity. For such models, the cooling curve hardly depends on the equation of state (EOS) as well as the mass of neutron star, while we show that the gravitational wave frequencies strongly depend on the both properties. Then, we find that the frequencies of the fundamental and the 1st pressure mode multiplied with the stellar mass are well expressed as a function of the stellar compactness almost independently of the EOS. We also find that the frequency of the 1st gravity mode in later phase of the thermal evolution is strongly correlated with the stellar compactness. In addition, we derive the empirical formula estimating the threshold mass for the onset of the direct Urca inside the neutron star as a function of the nuclear saturation parameter. This formula will give us a constraint on the neutron star properties, if it would be observationally found that the direct Urca occurs (or does not work) inside the neutron star.
\end{abstract}

%
\maketitle


\section{Introduction}
\label{sec:I}

A neutron star, which is a massive remnant of supernova explosion, is one of the most suitable object for probing the physics under extreme conditions. It is generally considered that the density inside the star significantly exceeds the standard nuclear density and the gravitational and magnetic fields can become much stronger than those observed in our solar system \cite{ST83}. The observations of the neutron star itself and/or the phenomena associated with the neutron star tell us an aspect of such extreme conditions. In practice, the discovery of the $2M_\odot$ neutron stars \cite{D10,A13,C20} has ruled out some of the soft equations of state (EOSs), with which the maximum mass does not reach the observed mass. Meanwhile, it simultaneously brings up the so-called hyperon puzzle, i.e., it becomes quite difficult to construct the EOS with hyperons, which overcomes the $2M_\odot$ neutron star observations. The gravitational waves from the binary neutron star merger, GW170817, also tell us the constraint on the $1.4M_\odot$ neutron star radius, i.e., $R_{1.4}\le 13.6$ km \cite{Annala18}, through the constraint on the tidal deformability. Moreover, the light bending, which is one of the relativistic effects produced by the strong gravitational field induced by the compact object, can change the pulsar light curve, which mainly tells us the constraint on the stellar compactness, $M/R$, with the stellar mass $M$ and radius $R$ (e.g., Refs. \cite{PFC83,LL95,PG03,PO14,SM18,Sotani20a}). Actually, the Neutron star Interior Composition Explorer (NICER), which is installed on an International Space Station (ISS), has already reported the constraint on the neutron star mass and radius for PSR J0030+0451 \cite{Riley19,Miller19} and for PSR J0740+6620 \cite{Riley21,Miller21}.

In addition to the current constraints on the neutron star mass and radius, the oscillation frequencies (and also gravitational waves) from the compact objects must become another information for seeing the properties of the source objects, if they will be observed. In practice, one would extract the neutron star properties as an inverse problem via the observation of the stellar oscillations, because their frequencies (and damping rate) strongly depend on the interior information of the neutron star. This technique is known as asteroseismology. For example, by identifying the quasiperiodic oscillations observed in the afterglow following the giant flares with the crustal torsional oscillations, the crustal EOS has been constrained (e.g., Refs. \cite{GNHL2011,SNIO2012,SIO2016}). In a  similar way, if one would observe the gravitational waves and identify it with a specific oscillation mode, one could extract the neutron star mass, radius, and EOS (e.g., \cite{AK1996,AK1998,STM2001,SH2003,TL2005,SYMT2011,PA2012,DGKK2013,Sotani20b,Sotani21}). Moreover, in order to understand the physical background behind the gravitational wave signals appearing in the numerical simulation data for core-collapse supernovae, the asteroseismology is becoming one of the important studies (e.g., \cite{FMP2003,FKAO2015,ST2016,SKTK2017,MRBV2018,SKTK2019,TCPOF19,SS2019,ST2020,STT2021}).

Nevertheless, the study about the gravitational wave asteroseismology on the neutron stars during thermal evolution is quite few (e.g., \cite{KHA15}). After the neutron star is born through the supernova explosion at the end of a massive star, it keeps on cooling due to the emissions of neutrinos ($t\lesssim10^5~{\rm yrs}$) and photons ($t\gtrsim10^5~{\rm yrs}$). During the cooling phase, the matter inside the neutron star is in the beta equilibrium state, where one has to also care the thermal profiles inside the star. Especially for $t\lesssim10^{2}~{\rm yrs}$, the thermal profile is generally not isothermal, where the neutron star core is much cooled compared with other outside regions. In such a period, the neutrino cooling in the crust highly contributes to the surface temperature~(e.g., Ref.~\cite{Page2020}). After $t\sim 10^{2}~{\rm yrs}$, the heat in the core is almost transported to the surface~\cite{Lattimer1994,Gnedin2001,Sales2020}, where the neutrino cooling in the core dominates thermal evolution of neutron stars. For $t\gtrsim10^5~{\rm yrs}$, the photon cooling becomes dominant. The calculated cooling curves can be checked by cooling observations including the age and surface temperature. Up to now, many studies for the neutron star cooling have been performed with various models, in order to extract the uncertain information of neutron stars, such as EOS, stellar mass, atmosphere composition, and the effect of nucleon superfluidity/superconductivity~(for review, see Refs.~\cite{YP04,Page06,Burgio2021}). Among them, since the superfluidity/superconductivity is still quite uncertain and strongly depends on the theoretical models, we simply neglect the effect of superfluidity/superconductivity in this study,
even though at least Cassiopeia A observation might be considered as the evidence of neutrons superfluidity \cite{Page2011,Shternin2011}.
On the other hand, we focus on the neutron star models with different EOSs and stellar mass without the accretion from the companion, i.e., the atmosphere is composed of only Ni. In this situation, the cooling curves hardly depend on the EOS and stellar mass (as shown in Fig. \ref{fig:Ts}), if the rapid cooling process due to the direct Urca does not work. That is, only by observing the cooling curves, it is quite difficult to distinguish the EOS and stellar mass. However, if the gravitational wave from such neutron star models would be observed, one may extract some of the neutron star properties. Moreover, many cooling observations can be well explained by {\it minimal cooling scenario}, which excludes any rapid cooling process~\cite{Page2004}. So, in this study, we focus on the neutron star models, where the direct Urca does not work, and consider the possibility for extracting the properties on the cooling neutron stars via gravitational wave observations. 
%

This manuscript is organized as follows. In Sec. \ref{sec:EOS} we show the models of cooling neutron stars, which become the background models for the linear analysis. In Sec. \ref{sec:Oscillation} we discuss the dependence of the gravitational wave frequencies on the neutron star properties during the cooling evolution. Finally, in Sec. \ref{sec:Conclusion} we conclude this study. Unless otherwise mentioned, we adopt geometric units in the following, $c=G=1$, where $c$ denotes the speed of light, and the metric signature is $(-,+,+,+)$.

\section{Model of cooling neutron stars}
\label{sec:EOS}

\begin{figure}[tbp]
\begin{center}
\includegraphics[scale=0.6]{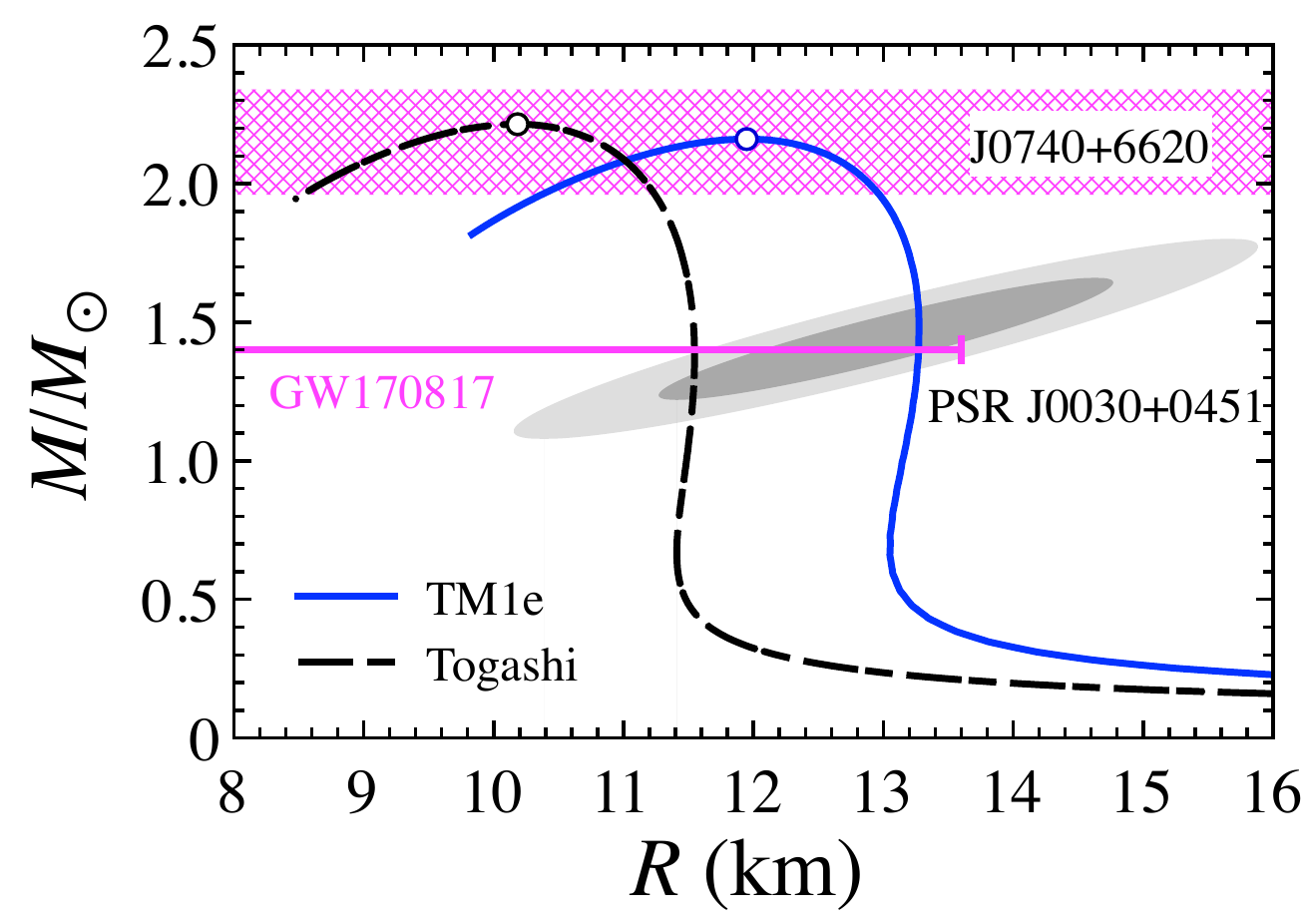} 
\end{center}
\caption{
Neutron star mass and radius relations constructed with the EOSs adopted in this study. The circles denote the neutron star models with the maximum mass expected with the corresponding EOS. For reference, we also show the constraint on the $1.4M_\odot$ neutron star radius derived by the GW170817 event, i.e., $R_{1.4}\le 13.6$ km \cite{Annala18}, the maximum mass observed so far, i.e., $M= 2.14^{+0.20}_{-0.18} M_\odot$ for MSP J0740+6620 \cite{C20}, and the constraint on the mass and radius derived by the NICER observation for PSR J0030+0451 \cite{Riley19,Miller19}. The inner and outer region of the NICER constraint denote the $1\sigma$ and $2\sigma$ levels \cite{Blaschke20}. 
}
\label{fig:MR}
\end{figure}

First, we have to prepare the background neutron star models for doing the linear analysis. In this study, we especially focus on the cooling neutron stars. In order to see the EOS dependence, we adopt two different EOSs, i.e., TM1e \cite{TM1e} and Togashi \cite{Togashi17}. TM1e is constructed in the relativistic mean field framework with the same nuclear interaction (TM1) adopted in the Shen EOS \cite{Shen}, but the interaction of isovector is changed in such a way that the density-dependence of symmetry energy, $L$, at the saturation point is adjusted. Togashi is the EOS based on the variational many-body theory with AV18 two-body potential and UIX three-body potential. Their EOS parameters, the maximum mass for the cold neutron star constructed with each EOS, and the threshold mass for the onset of the direct Urca, $M_{\rm DU}/M_\odot$, are listed in Table \ref{tab:EOS}, where $K_0$ is the incompressibility of the symmetric nuclear matter. Meanwhile, $\eta$ is a specific combination of $K_0$ and $L$ as $\eta\equiv (K_0 L^2)^{1/3}$ \cite{SIOO14}, which is a good property for expressing the low-mass neutron stars \cite{SSB16} and for discussing the neutron star maximum mass \cite{Sotani17,SK17}. Considering the constraint obtained from the GW170817, i.e., the $1.4M_\odot$ neutron star radius is less than 13.6 km \cite{Annala18}, $\eta$ may be constrained as $\eta \lesssim 125$ MeV. In Fig. \ref{fig:MR}, we show the neutron star mass and radius relations constructed with two EOSs, where we also show a couple of constraints obtained from the astronomical observations. From this figure, one can see that the EOSs adopted here overcome at least these astronomical constraints. In addition, to see the neutron star mass dependence, we adopt the neutron star models with the mass listed in Table \ref{tab:Baryon_mass}, where $M_{0}$ and $M_{\rm B}$ denote the initial gravitational mass and the baryon mass, respectively. Hereafter, we identify the neutron star model with the combination of the selected EOS and initial gravitational mass, such as Togashi (2.0) for the neutron star model constructed with the Togashi EOS, whose initial gravitational mass is $2.0M_\odot$.

\begin{table}
\caption{EOS parameters adopted in this study, the maximum mass, $M_{\rm max}/M_\odot$, for the cold neutron star constructed with each EOS, and the threshold mass for the onset of the direct Urca, $M_{\rm DU}/M_\odot$. The EOS dependence in the cooling curve and gravitational wave frequency is discussed with TM1e and Togashi EOSs, while the other EOSs are considered just for discussion about the EOS dependence of $M_{\rm DU}$ in Fig.~\ref{fig:MDU}.
We note that the direct Urca process does not turn on inside the neutron stars constructed with some of EOSs listed here, because the maximum mass does not reach $M_{\rm DU}$.} 
\label{tab:EOS}
\begin {center}
\begin{tabular}{cccccc}
\hline\hline
EOS & $K_0$ (MeV) & $L$ (MeV) & $\eta$ (MeV) & $M_{\rm max}/M_\odot$ & $M_{\rm DU}/M_\odot$   \\
\hline
TM1e & 281 & 40.0  &  76.6 & 2.12 & 2.06  \\
Togashi & 245 & 38.7 & 71.6 & 2.21 & ---  \\
\hline
KDE0v & 229 & 45.2 & 77.6 & 1.97 & --- \\
SLy4 & 230 & 45.9 & 78.5 & 2.05 & --- \\
SLy2 & 230 & 47.5 & 80.3 & 2.06 & 2.06 \\
SLy9 & 230 & 54.9 & 88.4 & 2.16 & 1.72 \\
LS220 & 220 & 73.8 & 106 & 2.06 & 1.35 \\
SkMp & 231 & 70.3 & 105 & 2.11 & 1.32 \\
SKa & 263 & 74.6 & 114 & 2.22 & 1.23 \\
Miyatsu  & 274 & 77.1 & 118 & 1.95 & 1.25  \\
SkI3 & 258 & 101 & 138 & 2.25 & 0.92 \\
Shen & 281 & 111 & 151 & 2.17 & 0.77 \\
\hline \hline
\end{tabular}
\end {center}
\end{table}

\begin{table}
\caption{Baryon mass, $M_{\rm B}$, for the neutron star models with various initial gravitational mass, $M_{0}$. } 
\label{tab:Baryon_mass}
\begin {center}
\begin{tabular}{ccc}
\hline\hline
EOS & $M_{0}/M_\odot$ & $M_{\rm B}/M_\odot$    \\
\hline
TM1e & 1.1 &  1.224   \\
          & 1.4  &  1.586   \\
          & 1.7  &  1.962   \\
          & 2.0  &  2.360   \\
          & 2.1  &  2.464   \\
Togashi & 1.1 & 1.206   \\
          & 1.4  & 1.573    \\
          & 1.7  & 1.968    \\
          & 2.0  &  2.391   \\
          & 2.1  &  2.534   \\
\hline \hline
\end{tabular}
\end {center}
\end{table}

After the EOS and the neutron star mass are selected, how to calculate the neutron star cooling evolution is basically the same as in Refs. \cite{Fujimoto84,Dohi19}. In this study, we simply consider the spherically symmetric neutron stars, whose metric is given by 
\begin{equation}
  ds^2 = -e^{2\Phi}dt^2 + e^{2\Lambda}dr^2 + r^2(d\theta^2 + \sin^2\theta d \phi^2), \label{metric}
\end{equation}
where the metric functions, $\Phi$ and $\Lambda$, are the function of $r$, and $\Lambda$ is directly associated with the mass function, $m(r)$, via $e^{-2\Lambda}=1-2m/r$. The thermal evolution inside the neutron star is calculated via the following equations, together with the Tolman-Oppenheimer-Volkoff equations;
\begin{gather}
  \frac{e^{-\Lambda-2\Phi}}{4\pi r^2}\frac{\partial}{\partial r}\left(e^{2\Phi} L_r\right) 
       = -Q - \frac{C_v}{e^{\Phi}}\frac{\partial T}{\partial t}, \label{eq:eq2}\\
  \frac{L_r}{4\pi \kappa r^2} = e^{-\Lambda-\Phi}\frac{\partial}{\partial r}\left(Te^{\Phi}\right),
\end{gather}
where $T$, $L_r$, $C_v$, $\kappa$, and $Q$ are the local temperature, the local luminosity, the heat capacity per unit volume, thermal conductivity, and the neutrino emissivity, respectively \cite{YP04}. We note that these equations are practically integrated with the mass coordinate instead of the radial coordinate $r$ (see Refs. \cite{Fujimoto84,Dohi19,Dohi21} for the concrete equations), in order to carefully treat the region in the vicinity of the stellar surface. Specifically, we impose the radiative zero boundary condition at sufficiently closed area to the photosphere on the surface~\cite{Fujimoto84}. We obtain the specific heat $C_V$ from the linear interpolation of public data of the Togashi and TM1e. For the opacity $\kappa$, we consider the radiative opacity~\cite{Schatz1999} and conductive opacity composed of mainly electrons~\cite{Potekhin2015} and neutrons~\cite{Baiko2001}. 
In addition, since in this study we neglect the effect of superfluidity/superconductivity in the neutron star cooling, the main neutrino emission processes, which are included in $Q$ in Eq.~(\ref{eq:eq2}), are as follows;
\begin{gather}
   \begin{cases} 
        n+n^\prime \rightarrow p+n^\prime+l^-+\bar\nu_l, \\ 
        p+n^\prime+l^- \rightarrow n+n^\prime+\nu_l, 
   \end{cases}    \label{eq:MUn} \\
   \begin{cases} 
       n+p^\prime \rightarrow p+p^\prime+l^-+\bar\nu_l, \\ 
       p+p^\prime+l^- \rightarrow n+p^\prime+\nu_l, 
   \end{cases}   \label{eq:MUp} \\
   \begin{cases} 
       n+n^\prime \rightarrow n+n^\prime+\nu_{l}+\bar\nu_{l}, \\ 
       n+p \rightarrow n+p+\nu_{l}+\bar\nu_{l},  \\
       p+p^\prime \rightarrow p+p^\prime+\nu_{l}+\bar\nu_{l}, 
    \end{cases}  \label{eq:Bre} \\
    \begin{cases} 
        n \rightarrow p+l^-+\bar\nu_l, \\ 
        p+l^- \rightarrow n+\nu_l,
   \end{cases}   \label{eq:DU}
\end{gather}
where $l$ denotes the electron or muon. The processes expressed with Eqs. (\ref{eq:MUn}) -- (\ref{eq:DU}) correspond to the modified Urca (neutron branch), the modified Urca (proton branch), the nucleon pair bremsstrahlung, and the nucleon direct Urca, respectively. The modified Urca and bremsstrahlung processes are slow cooling processes and always turn on, where the energy loss rate is $\sim 10^{10-21}$ erg/cm$^3$/s \cite{ST83}. We also consider the electron-positron pair creation, photo-neutrino process, and plasmon decay processes, which are much weaker than the above cooling processes. On the other hand, the direct Urca process is rapid cooling process, whose emissivity is $\sim 10^{27}$ erg/cm$^3$/s \cite{LPPH91}, but this process does not always turn on due to the momentum conservation, as discussed below.
We note that the EOS adopted in this study does not include the presence of hyperons, which may cause additional cooling channels such as the $\Lambda$-hyperons Direct Urca process~\cite{Prakash92,Tsuruta09,Raduta18}. 

\begin{figure}[tbp]
\begin{center}
\includegraphics[scale=0.6]{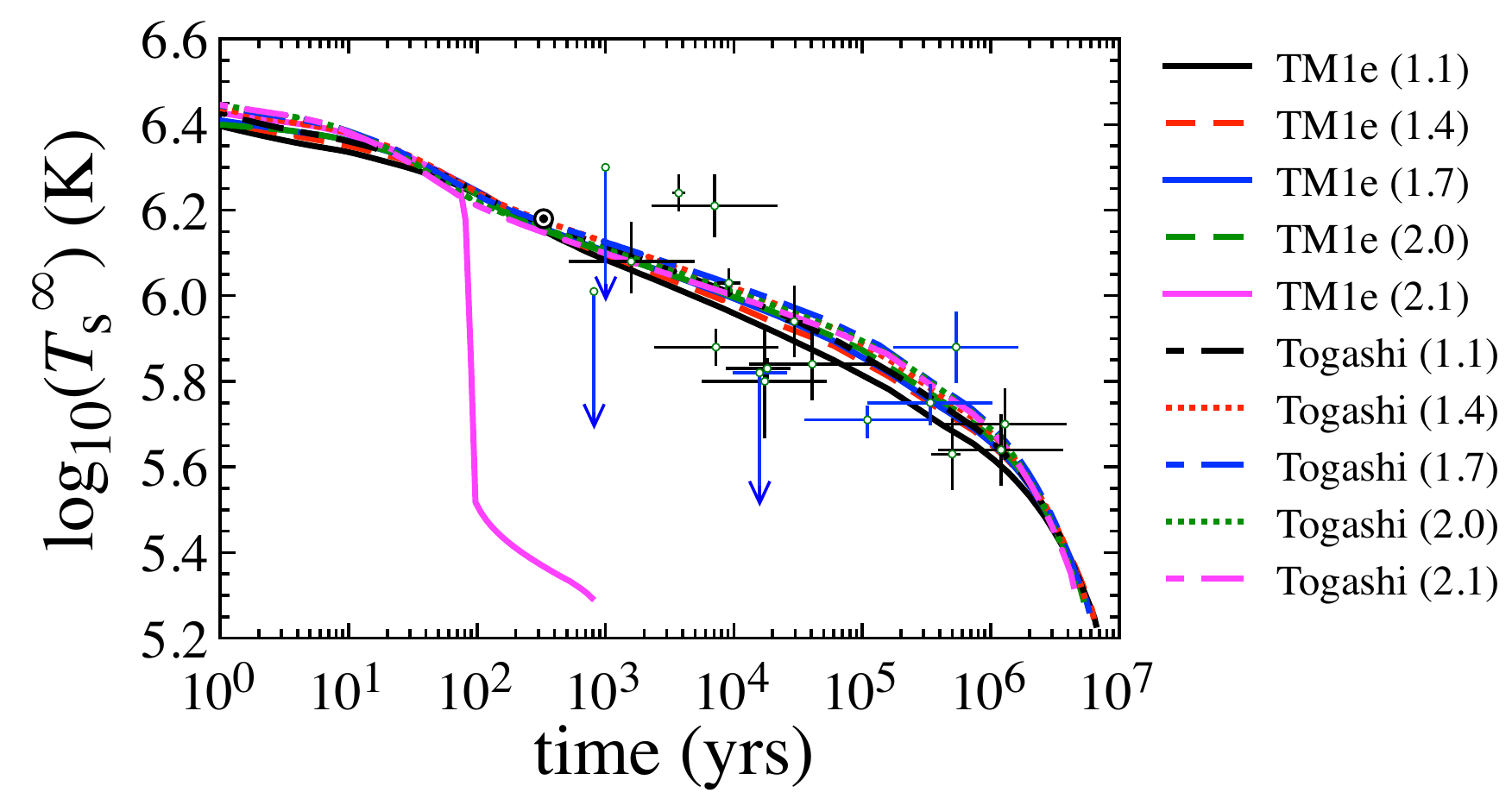} 
\end{center}
\caption{
Cooling curves calculated for the neutron stars with TM1e and Togashi, where $T_{\rm s}^\infty$ denotes the effective surface temperature of neutron star observed at infinity. One can observe the rapid cooling only for the $2.1M_\odot$ neutron star model constructed with TM1e among the neutron star models considered in this study. For reference, we also show the cooling data for Cassiopeia A (double circle) and the other objects (circle with error) \cite{LHL17}.
}
\label{fig:Ts}
\end{figure}

\begin{figure}[tbp]
\begin{center}
\includegraphics[scale=0.6]{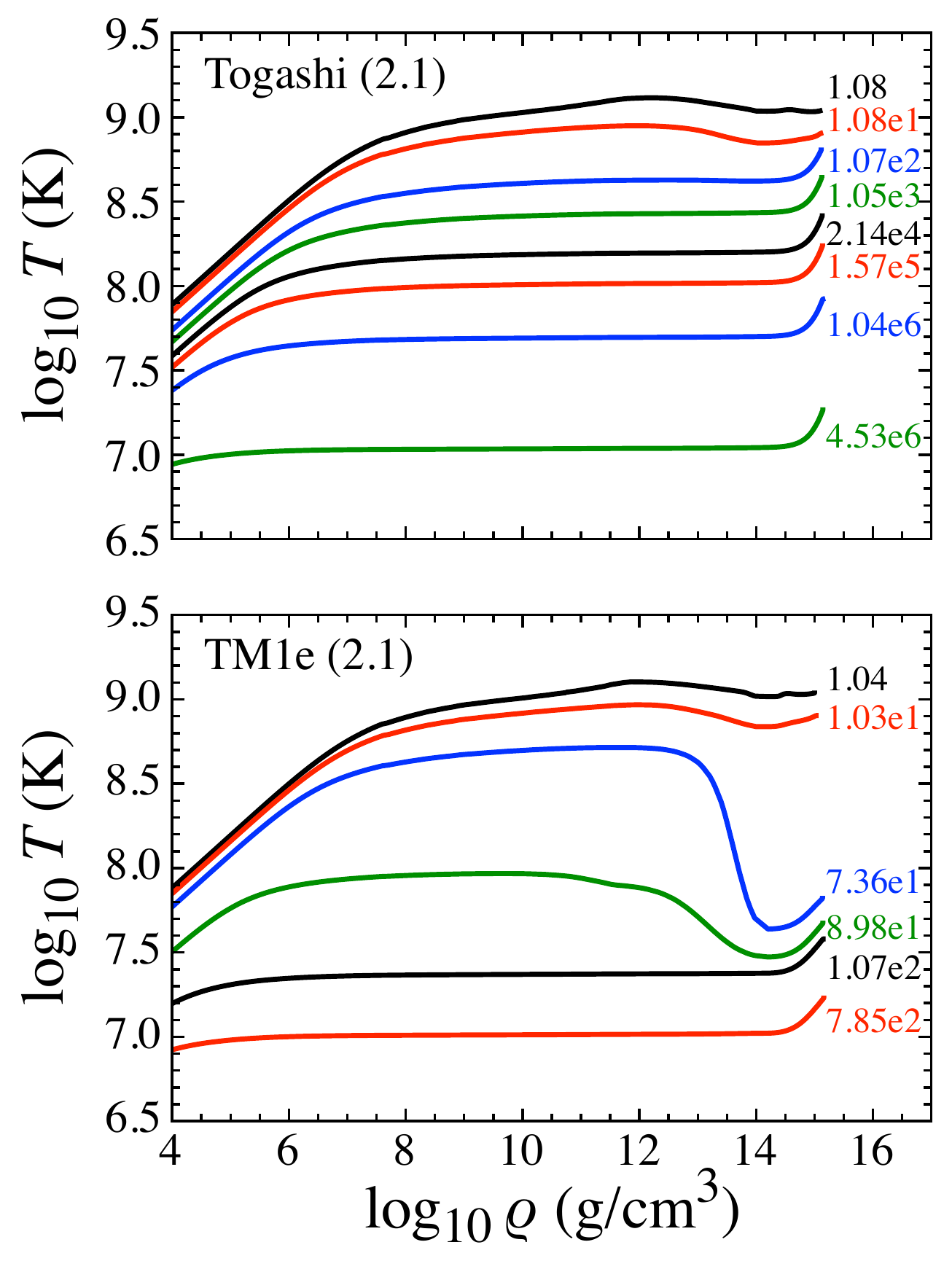} 
\end{center}
\caption{
The thermal distribution inside the neutron star for the models of Togashi (2.1) and TM1e (2.1) with different time. In each panel, the labels next to the lines denote the time after the neutron star is born in the unit of years. 
}
\label{fig:Trho-21}
\end{figure}

Then, we assume the isothermal condition at $t=0$, i.e., especially in this study we adopt the condition that $Te^\Phi\simeq10^{9.3}~{\rm K}$ inside the neutron star. 
The initial temperature may change the thermal evolution. But, the dependence on the initial temperature seems to appear only up to $\sim 10^4$ sec, the so-call neo-neutron stars era \cite{BPR20}. In this study, to avoid this dependence, we only consider the neutron star models at more than 1 year after they are born.
The resultant cooling curves for various neutron star models are shown in Fig. \ref{fig:Ts}, where for reference we also show some of the observed cooling data. In this figure, one can observe the rapid cooling only for the model of TM1e (2.1) among the neutron star models considered here. One can also see that the dependence of the cooling curve on the EOS and initial mass is quite weak, if the rapid cooling processes do not turn on. Namely, only by observing the thermal evolution, it is quite difficult to distinguish the EOS and mass of the neutron star, in which the rapid cooling does not work. So, in this study, focusing only on the neutron star models where the rapid cooling does not work, we consider the possibility for extracting the properties via the gravitational wave observations from such  objects. We note that the atmosphere is assumed in this study to be composed of only nickel (Ni), i.e., without any accreting matter. But, anyway if one assumes specific compositions in the atmosphere, again the cooling curve hardly depends on the EOS and initial mass of neutron star as in Fig. \ref{fig:Ts} (e.g., \cite{Page06,LHL17}).
Additionally, in Fig. \ref{fig:Trho-21} we show the thermal profile inside the neutron stars for Togashi (2.1) and TM1e (2.1) as an example of the evolution of the thermal profile without and with the direct Urca process. One can observe that the shape of the thermal profile is almost keeping without the direct Urca process, while it dramatically changes during the rapid cooling process turns on. 
We note that if one takes into account the superfluity/superconductivity effects, the additional cooling process due to the pair breaking and formation may turn on, which modifies the cooling curve depending on the stellar mass (e.g., \cite{Gusakov05,Shternin11}). In addition, if one considers this cooling process together with the direct Urca process, the corresponding cooling curves become more complex, which may appear the region between the curves without and with the direct Urca process in Fig. \ref{fig:Ts}, e.g., Fig. 4 in Ref. \cite{Gusakov05}.

\begin{figure*}[tbp]
\begin{center}
\includegraphics[scale=0.6]{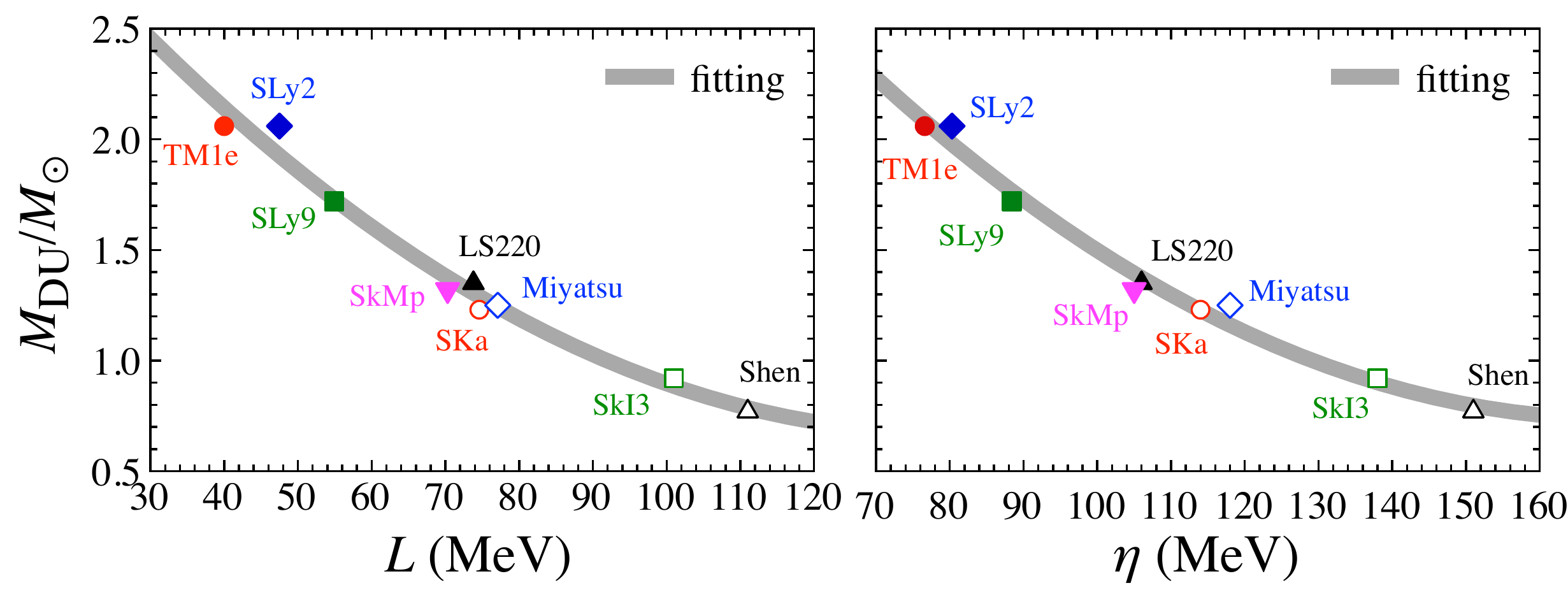} 
\end{center}
\caption{
The threshold mass for the onset of the direct Urca is shown as a function of $L$ (left panel) and $\eta$ (right panel) for various EOS models listed in Table~\ref{tab:EOS}. The thick-solid lines denote the fitting formulas given by Eqs. (\ref{eq:MDR1}) and (\ref{eq:MDR2}). The direct Urca process basically occurs in the region above these lines.  
}
\label{fig:MDU}
\end{figure*}

The condition for the onset of direct Urca is expressed as $Y_p\ge Y_p^{\rm DU}$ \cite{LPPH91}, where $Y_p$ is the proton fraction and $Y_{p}^{\rm DU}$ is the critical value of $Y_p$ given by 
\begin{equation}
  Y_p^{\rm DU} = \left[1+ \left(1+\chi_e^{1/3}\right)^3\right]^{-1}, \label{eq:YpDU}
\end{equation}
where $\chi_e=Y_e/(Y_e+Y_\mu)$ with the electron and muon fraction, $Y_e$ and $Y_\mu$. Considering that $Y_e\simeq Y_\mu$ for a high density region, $Y_p^{\rm DU}\simeq 0.1477$, while $Y_p^{\rm DU}=1/9$ if muons are absent. The direct Urca process occurs for more massive stars, because $Y_p$ generally increases with density in a high density region. In addition, $Y_p$ strongly depends on the nuclear symmetry energy through $L$, which in turn must depend on $\eta$. In practice, as shown in Fig. \ref{fig:MDU}, we find the threshold mass for the onset of the direct Urca, which is listed in Table \ref{tab:EOS}, is strongly correlated with $L$ or $\eta$, with which we can derive the fitting formulas for such a threshold mass as a function of $L$ or $\eta$;
\begin{gather}
  M_{\rm DU}/M_\odot = 3.5801 -4.2036 L_{100} +  1.5191 L_{100}^2,  \label{eq:MDR1} \\
  M_{\rm DU}/M_\odot = 5.1515 -5.1629\eta_{100} +  1.5091\eta_{100}^2,  \label{eq:MDR2}
\end{gather}
where $L_{100}$ and $\eta_{100}$ are given by $L_{100}\equiv L/(100\ {\rm MeV})$ and $\eta_{100}\equiv \eta/(100\ {\rm MeV})$. The direct Urca process basically occurs inside the neutron stars with the mass above the fitting line. That is, this fitting will give us a constraint of the neutron star properties, i.e., the stellar mass and $L$ (or $\eta$) of the EOS must be in the region above the fitting line in Fig. \ref{fig:MDU}, if it would be observationally found that the direct Urca occurs inside the neutron star. We note that the direct Urca process does not turn on inside the neutron stars constructed with some of EOSs listed in Table \ref{tab:EOS}, because the maximum mass does not reach $M_{\rm DU}$ (or the central density for the neutron star with the maximum mass is still lower than the critical density, with which $Y_p= Y_p^{\rm DU}$).
Anyway, the neutron star models considered in this study are  basically in the region below the fitting lines given by Eqs. (\ref{eq:MDR1}) and~(\ref{eq:MDR2}).

\begin{figure*}[tbp]
\begin{center}
\includegraphics[scale=0.5]{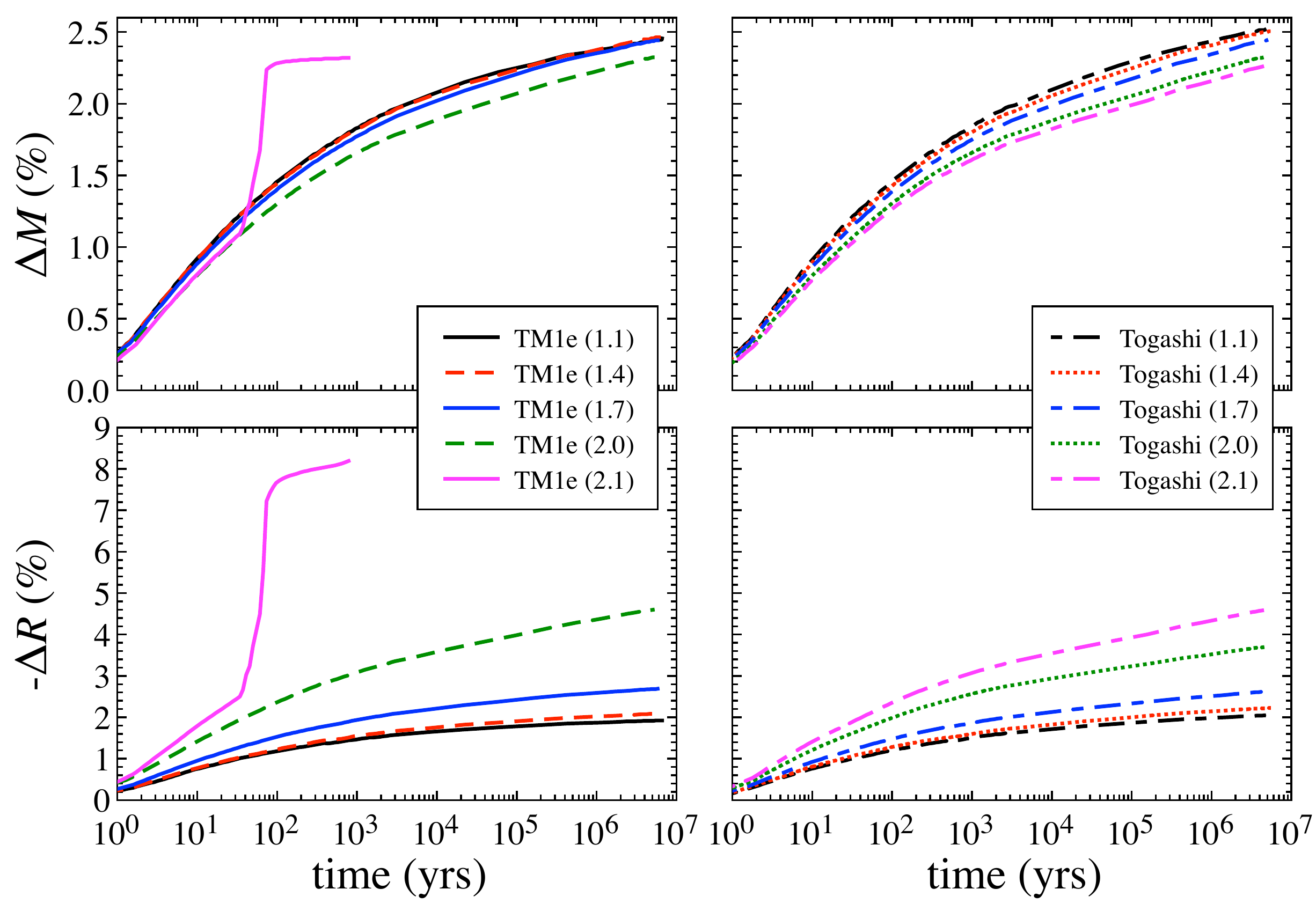} 
\end{center}
\caption{
Relative changes in the gravitational mass ($\Delta M$) and radius ($\Delta R$) are shown as a function of time, where $\Delta M$ and $\Delta R$ are calculated with Eqs. (\ref{eq:DM}) and (\ref{eq:DR}). The left and right panels correspond to the results for the neutron star models with TM1e and Togashi, respectively. 
}
\label{fig:DMR}
\end{figure*}

In this study, we consider the neutron star model whose baryon mass is constant during the thermal evolution, where the stellar radius is shrinking and the gravitational mass is increasing a little, because the thermal pressure is decreasing due to the neutrino cooling. In practice, in Fig. \ref{fig:DMR} we show the changes in the gravitational mass and radius calculated with
\begin{gather}
  \Delta M  = (M(t) - M_{0})/M_{0}, \label{eq:DM} \\
  \Delta R  = (R(t) - R_{0})/R_{0}, \label{eq:DR}
\end{gather}
where $M(t)$ and $R(t)$ denote the gravitational mass and the stellar radius at a specific time $t$, while $R_{0}$ is the initial radius. From this figure, one can observe that the gravitational mass increases a few percent of the initial gravitational mass, which is less dependence on the initial gravitational mass. On the other hand, the stellar radius shrinks $\sim 5\%$ of the initial radius for minimum cooling, while that shrinks more for rapid cooling. As a result, the stellar compactness, $M/R$, increases with time.

At the end of this section, we remark that the difference between the models discussed in Ref. \cite{KHA15} and ours. In Ref. \cite{KHA15}, the crustal elasticity is taken into account,  but they calculated the thermal profile up to the neutron star crust with zero-temperature EOS, adopting Fermi gas approximation for including thermal effect, and discussed the surface temperature determined 
with the so-called $T_s-T_b$ relation, which is analytical formula between envelope and crust surface temperature~\cite{Gudmundsson1983}. On the other hand, in this study we neglect the crustal elasticity, but the cooling calculations have been done up to the stellar surface with finite temperature EOS.

\section{Gravitational wave asteroseismology}
\label{sec:Oscillation}

In order to examine the gravitational wave frequencies, we conduct a linear analysis on the neutron star models discussed in the previous section, focusing only on the case where the direct Urca process does not work inside the star. In this study we simply adopt the relativistic Cowling approximation, where the metric perturbations are neglected during the fluid oscillations. Even with this simple approximation, one can qualitatively discuss the behavior of the gravitational wave frequency \cite{ST2020}. The concrete perturbation equations and the boundary conditions to be imposed are completely the same as in Ref. \cite{SKTK2019}.

\begin{figure}
\begin{center}
\includegraphics[scale=0.6]{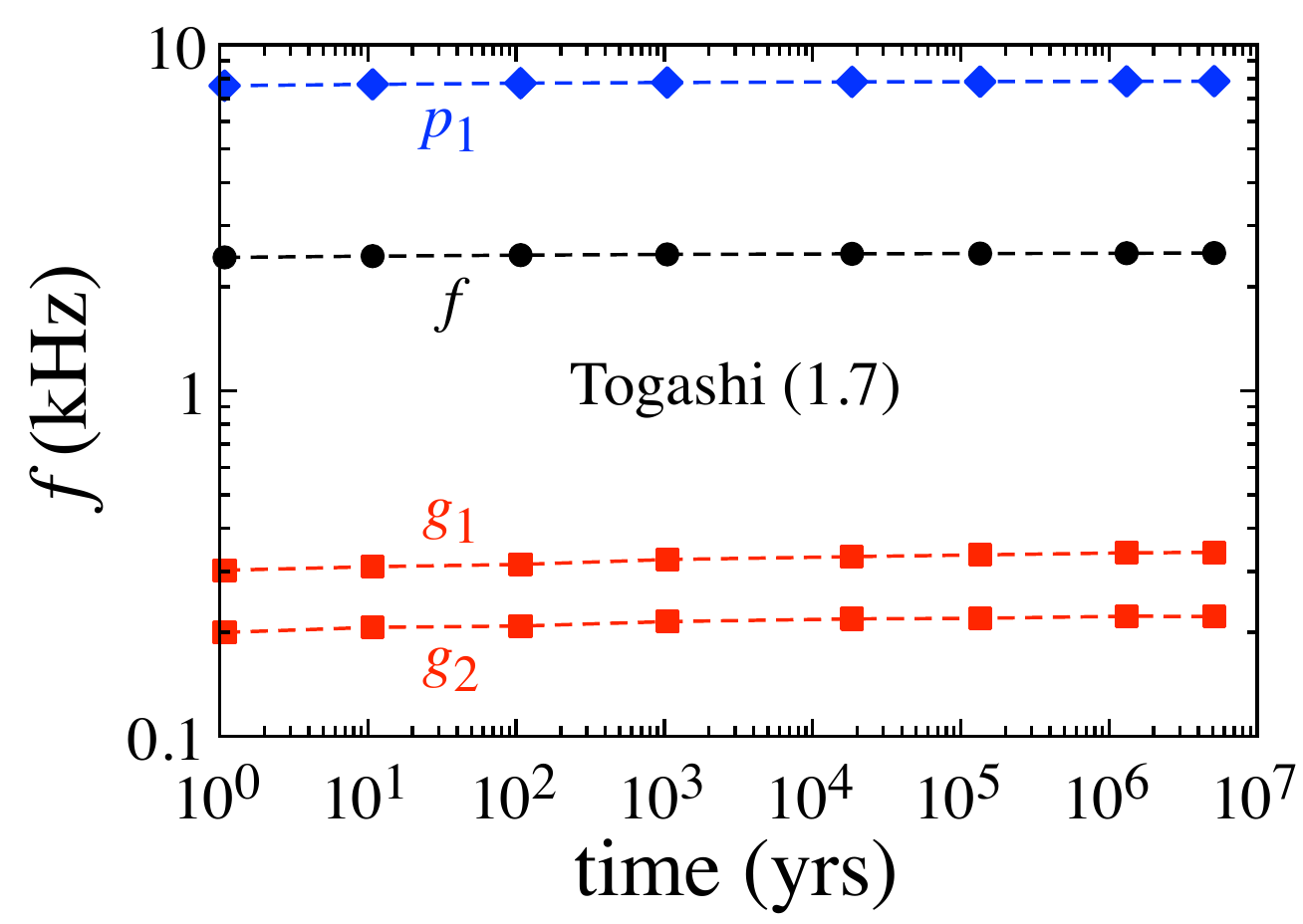} 
\end{center}
\caption{
The time evolution of the $f$-, $p_1$-, $g_1$-, and $g_2$-mode frequencies from the neutron star constructed with Togashi EOS, whose initial mass is $1.7M_\odot$.
}
\label{fig:f-TS-TG17}
\end{figure}

\begin{figure}
\begin{center}
\includegraphics[scale=0.5]{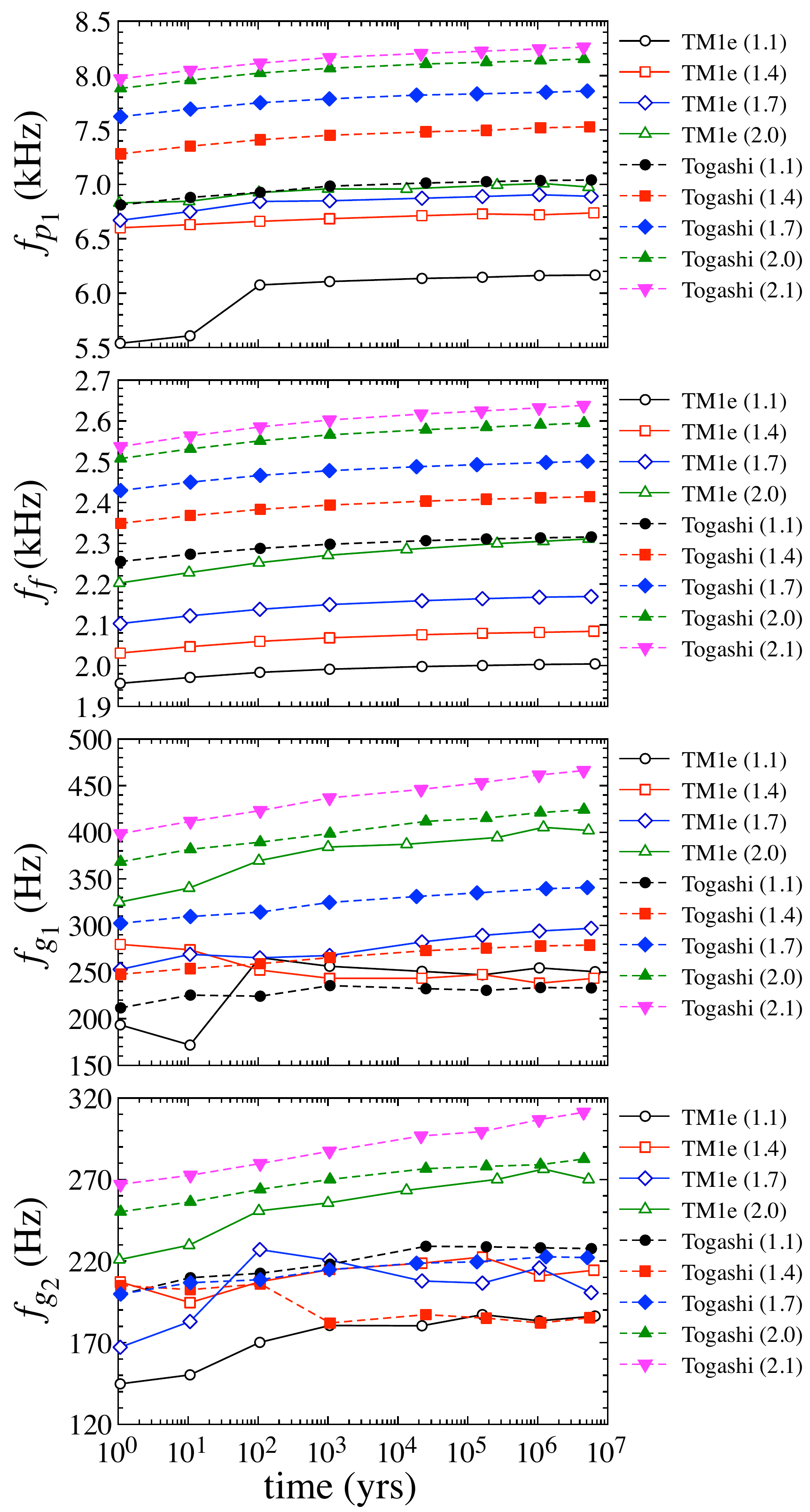} 
\end{center}
\caption{
The eigenfrequency evolution for various neutron star models. The panels from top to bottom correspond to the $p_1$-, $f$-, $g_1$-, and $g_2$-mode frequencies. 
}
\label{fig:ft}
\end{figure}

In Fig. \ref{fig:f-TS-TG17}, as an example we show the evolution of the gravitational wave frequencies from\ the model of Togashi (1.7), where we focus only on the fundamental ($f$), the 1st pressure ($p_1$), the 1st gravity ($g_1$), and the 2nd gravity ($g_2$) mode frequencies. From this figure, one can see that the frequency does not change dramatically at least for this model. In Fig. \ref{fig:ft}, we show the $p_1$-, $f$-, $g_1$-, and $g_2$-mode frequencies from top to bottom for various neutron star models. One can observe that the $f$- and $p_1$-mode frequencies are relatively quiet with time, while the $g$-mode frequencies are more dynamical especially for the neutron star models with lower mass and in the early phase. Anyway, one can see the dependence of the gravitational wave frequencies on the neutron star models, even though the rapid cooling does not work inside the star, where the cooling curve hardly depends on the models as mentioned above.

\begin{figure}
\begin{center}
\includegraphics[scale=0.6]{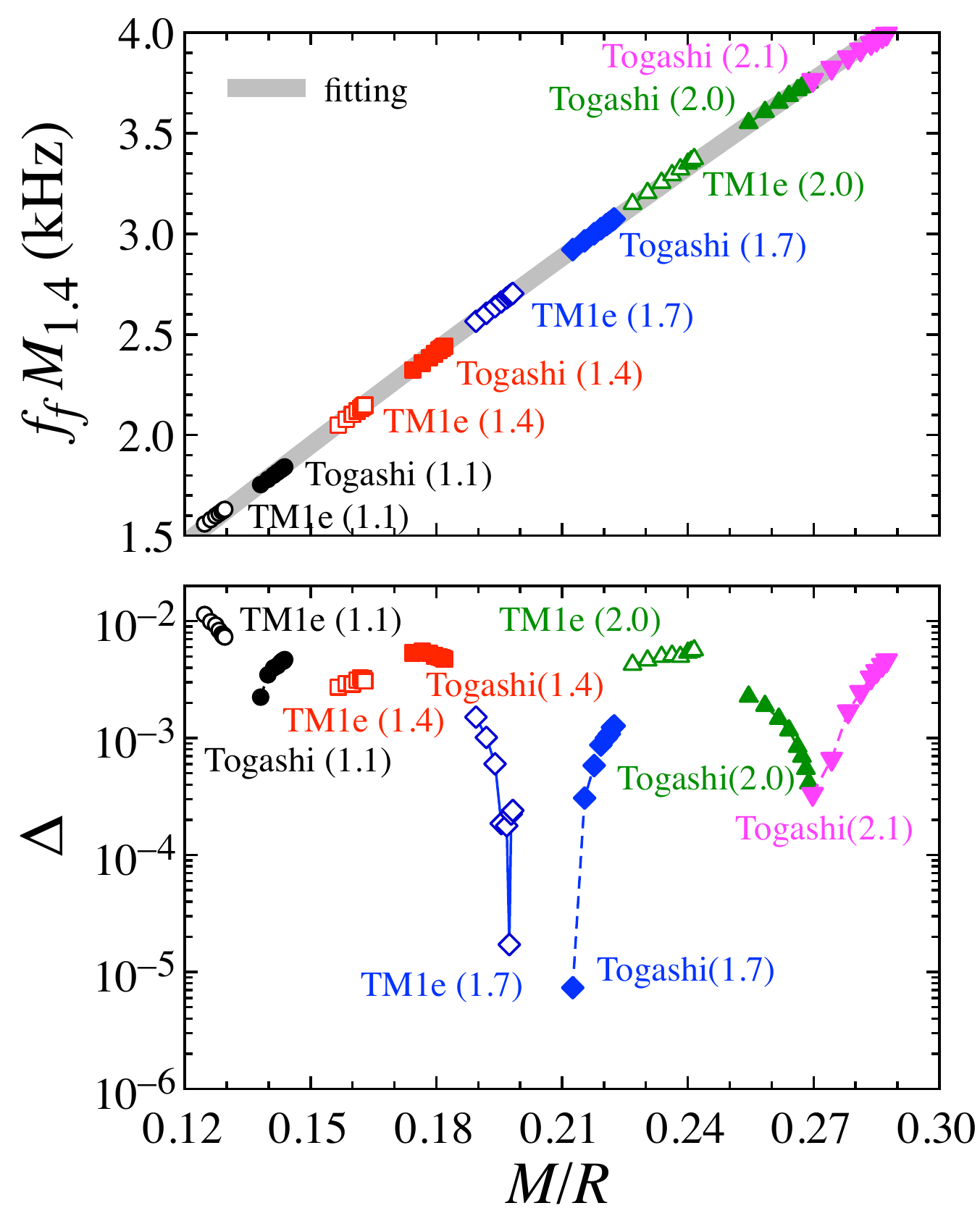} 
\end{center}
\caption{
In the top panel, the $f$-mode frequency multiplied with the normalized neutron star mass is plotted as a function of the stellar compactness for various models, where thick-solid line is a fitting given by Eq. (\ref{eq:ffM-comp}). Note that the open and filled marks correspond to the results for TM1e and Togashi, respectively. In the bottom panel, the relative deviation of the exact values calculated via eigenvalue problems from the values expected with the fitting formula (Eq. (\ref{eq:ffM-comp})).
}
\label{fig:ffM-comp}
\end{figure}

\begin{figure}[tbp]
\begin{center}
\includegraphics[scale=0.6]{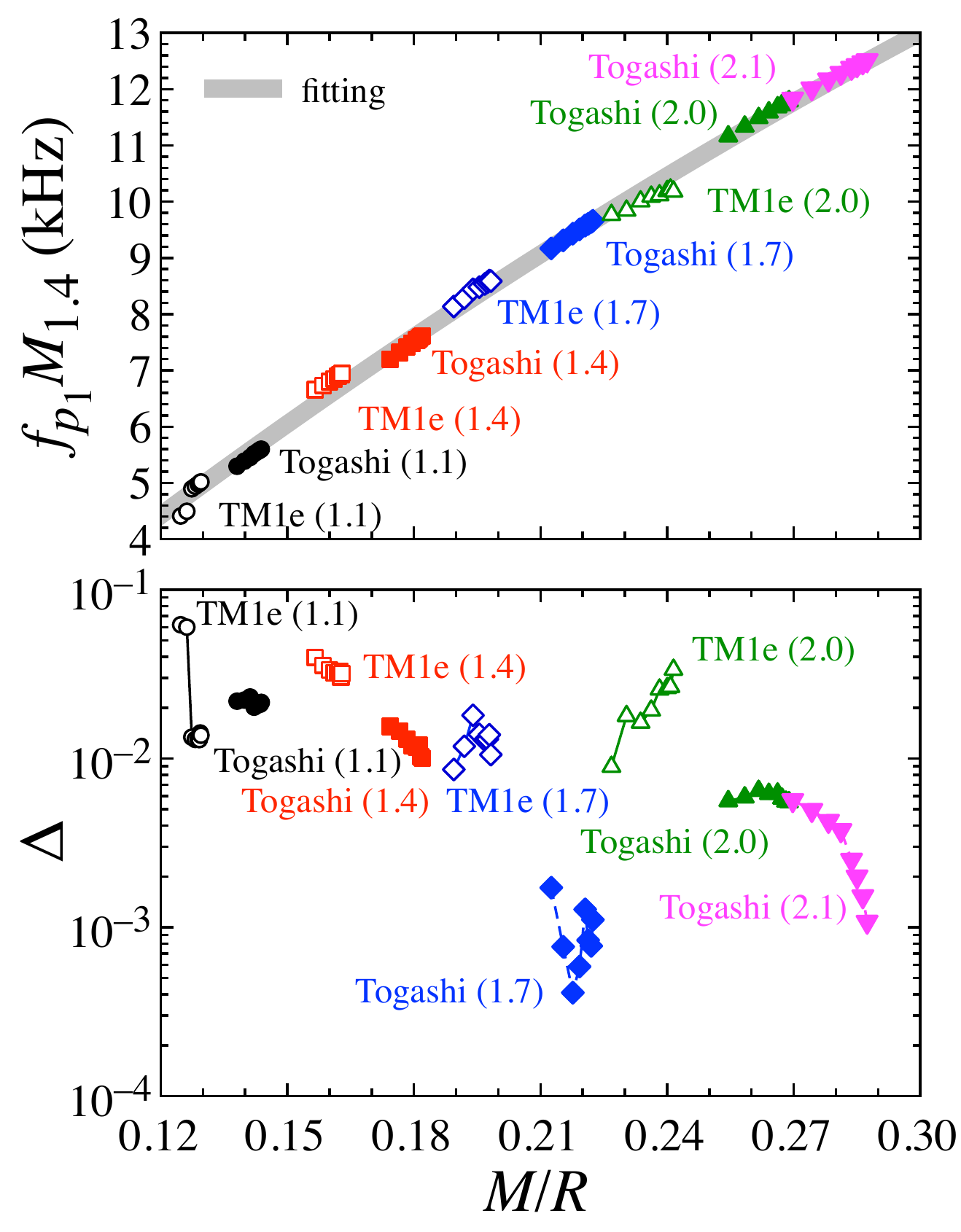} 
\end{center}
\caption{
Same as in Fig. \ref{fig:ffM-comp}, but for the $p_1$-mode frequency with the fitting given by Eq.(\ref{eq:fp1M-comp}).
}
\label{fig:fp1M-comp}
\end{figure}

In this situation, a kind of the empirical relation, which is a relation between the observables of neutron stars almost independently of the EOS, is quite useful for extracting the neutron star properties. In fact, for cold neutron stars, it is known that the $f$- and $p_1$-mode frequencies multiplied with the stellar mass are well expressed as a function of the stellar compactness \cite{TL2005,Sotani21}. In a similar way, we plot the same combination of the frequency and the normalized mass as a function of $M/R$ in the top panels of Figs. \ref{fig:ffM-comp} and \ref{fig:fp1M-comp}. From these figures, we  find that the $f$- and $p_1$-mode frequencies multiplied with the mass weakly depend on the neutron star models, and can derive the fitting formulas;
\begin{gather}
  f_f M_{1.4}\ ({\rm kHz}) = -0.6091 + 18.1198(M/R) - 7.0865(M/R)^2, \label{eq:ffM-comp} \\
  f_{p_1} M_{1.4}\ ({\rm kHz}) = -2.8393 + 65.6701(M/R) - 42.8130(M/R)^2, \label{eq:fp1M-comp}
\end{gather}
where $M_{1.4}$ denotes the stellar mass normalized $1.4M_\odot$, i.e., $M_{1.4}\equiv M/(1.4M_\odot)$. In the same figure, we also plot the prediction from the fitting formulas with the thick-solid lines, and show the relative deviation from the fitting formulas in the bottom panel. Thus, by detecting the $f$-mode ($p_1$-mode) frequency in the gravitational waves, one may draw one constraint in the mass and radius relation of neutron stars with $\sim 1\%$ ($\sim 10\%$) accuracy, even though one would not know the EOS and the neutron star mass.

\begin{figure}[tbp]
\begin{center}
\includegraphics[scale=0.6]{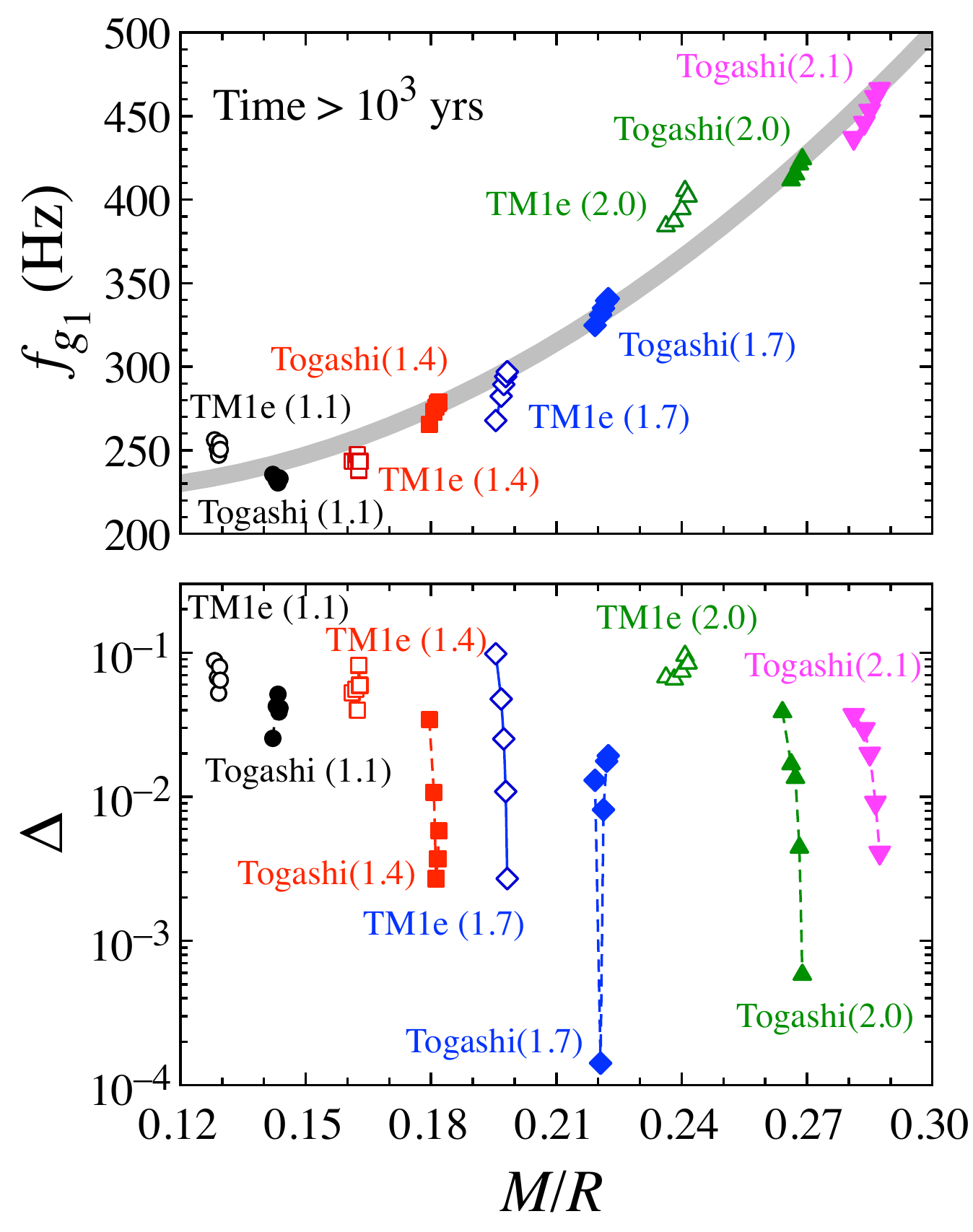} 
\end{center}
\caption{
For the neutron star models at more than $10^3$ yrs after they are born, the $g_1$-mode frequencies are shown as a function of stellar compactness together with the fitting formula (thick-solid line) given by Eq.(\ref{eq:g1}) in the top panel. The bottom panel denotes the relative deviation from the fitting formula for various neutron star models. 
}
\label{fig:fg1-comp}
\end{figure}

On the other hand, the $g$-mode frequencies seem to be more difficult to fit with a specific neutron star property. Nevertheless, focusing only on the neutron stars at more than $10^{3}$ yrs after they are born, i.e., in the later phase, we find a good correlation between the $g_1$-mode frequency and stellar compactness. In the top panel of Fig. \ref{fig:fg1-comp} we plot the $g_1$-mode frequency as a function of $M/R$ for various neutron star models, where the thick-solid line denotes the fitting formula given by
\begin{gather}
  f_{g_1} ({\rm Hz}) = 275.13 - 1124.6(M/R) + 6249.2 (M/R)^2. \label{eq:g1}
\end{gather}
In the bottom panel, we also show the relative deviation from the fitting formula, which tells us that one can predict the $g_1$-mode frequency as a function of $M/R$ within $\sim 10\%$ accuracy. Here, we note that this relation may be modified or this kind of relation may not exist, if one will take into account the effect of superfluidity and/or superconductivity in the thermal evolution, because the $g$-mode frequency may be more sensitive to the thermal profile inside the star, compared to the $f$- and $p$-mode frequencies. Thus, if the value of $M/R$ predicted with Eq. (\ref{eq:ffM-comp}) (or Eq. (\ref{eq:fp1M-comp})) is completely different from that with Eq. (\ref{eq:g1}), when one would detect the $f$-mode (or $p_1$-mode) frequency together with the $g_1$-mode frequency in gravitational waves from a  cooling neutron star, the superfluidity and/or superconductivity may play an important role inside the neutron star.

\section{Conclusion}
\label{sec:Conclusion}

Cooling curves, which are a result of the thermal evolution of neutron stars, are one of the important observables for seeing the neutron star properties. Nevertheless, without the effect of superfluidity and superconductivity, the cooling curve hardly depends on the EOS and mass of neutron star with a specific atmosphere model, if the direct Urca process does not turn on inside the star. In this study, we focus on such a neutron star model and consider to extract the neutron star property by observing gravitational waves, where the frequencies are determined by solving the eigenvalue problem with the relativistic Cowling approximation. Then, we find that the $f$- and $p_1$-mode frequencies multiplied with the stellar mass are well expressed as a function of the neutron star compactness independently of the EOS and mass of neutron stars. In a similar way, we also find that the $g_1$-mode frequencies in later phase of the thermal evolution is characterized by the compactness almost independently of the EOS and mass of neutron stars. Thus, using the resultant fitting relation, one can draw one constraint on the neutron star mass and radius relation via the direct observation of gravitational waves. Meanwhile, if the compactness predicted with the $f$-mode (or $p_1$-mode)  frequency would be completely different from that with the $g_1$-mode frequency, it may be an evidence that the superfluidity and/or superconductivity, which are neglected in this study, play an important role during the thermal evolution inside the neutron star. 

In addition, we find that the critical mass of the neutron star, with which the direct Urca process turns on, is strongly correlated with the density-dependence of the nuclear symmetry energy (or the suitable combination of the nuclear saturation parameter, $\eta$), and derive a kind of fitting formula. With using this relation, one would constrain the relation between the neutron star mass and nuclear saturation parameter, if one would observationally know that the direct Urca process occurs (or does not work) inside the neutron star. 

In this study, as a first step, we neglect the effect of superfluidity and/or superconductivity. We also neglect the effect of crust elasticity inside the neutron star, with which the additional oscillation modes, such as the shear ($s$) and interface ($i$) modes, could be excited as well as the $f$-, $p$-, and $g$-modes \cite{KHA15}. These effects will be taken into account somewhere in the future.

\acknowledgments

This work is supported in part by Japan Society for the Promotion of Science (JSPS) KAKENHI Grant Numbers 
JP19KK0354,  
JP20H04753,  and 
JP21H01088,  
and by Pioneering Program of RIKEN for Evolution of Matter in the Universe (r-EMU).



\begin{thebibliography}{999}

\bibitem{ST83}
   S. L. Shapiro and S. A. Teukolsky, {\it Black Holes, White Dwarfs, and Neutron Stars: The Physics of Compact Objects}  (Wiley-Interscience, New York, 1983).

\bibitem{D10} 
   P. Demorest, T. Pennucci, S. Ransom, M. Roberts, and J. Hessels, Nature {\bf 467}, 1081 (2010).

\bibitem{A13} 
   J. Antoniadis {\it et al.}, Science {\bf 340}, 6131 (2013).

\bibitem{C20}    
   H. T. Cromartie {\it et al.}, Nature Astronomy {\bf 4}, 72 (2020).

\bibitem{Annala18}  
   E. Annala, T. Gorda, A. Kurkela, and A. Vuorinen, Phys. Rev. Lett. {\bf 120}, 172703 (2018).

\bibitem{PFC83} 
   K. R. Pechenick, C. Ftaclas, and J. M. Cohen, Astrophys. J. {\bf 274}, 846 (1983).

\bibitem{LL95} 
   D. A. Leahy and L. Li, Mon. Not. R. Astron. Soc. {\bf 277}, 1177 (1995).

\bibitem{PG03} 
   J. Poutanen and M. Gierlinski, Mon. Not. R. Astron. Soc. {\bf 343}, 1301 (2003).

\bibitem{PO14} 
   D. Psaltis and F. \"{O}zel, Astrophys. J. {\bf 792}, 87 (2014). 

\bibitem{SM18} 
   H. Sotani and U. Miyamoto, Phys. Rev. D {\bf 98}, 044017 (2018); {\bf 98}, 103019 (2018).

\bibitem{Sotani20a} 
   H. Sotani, Phys. Rev. D {\bf 101}, 063013 (2020).

\bibitem{Riley19} 
   T. E. Riley {\it et al.}, Astrophys. J.  {\bf 887}, L21 (2019).
   
\bibitem{Miller19} 
   M. C. Miller {\it et al.}, Astrophys. J.  {\bf 887}, L24 (2019).
   
\bibitem{Riley21} 
   T. E. Riley {\it et al.}, Astrophys. J.  {\bf 918}, L27 (2021).
   
\bibitem{Miller21} 
   M. C. Miller {\it et al.}, Astrophys. J.  {\bf 918}, L28 (2021).
   
   

\bibitem{GNHL2011}
   M. Gearheart, W. G. Newton, J. Hooker, and B. -A. Li, Mon. Not. R. Astron. Soc. {\bf 418}, 2343 (2011).
   
\bibitem{SNIO2012}
   H. Sotani, K. Nakazato, K. Iida, and K. Oyamatsu, Phys. Rev. Lett. {\bf 108}, 201101 (2012);
   Mon. Not. R. Astron. Soc. {\bf 428}, L21 (2013); {\bf 434}, 2060 (2013).

\bibitem{SIO2016}
   H. Sotani, K. Iida, and K. Oyamatsu, New Astron. {\bf 43}, 80 (2016);
   Mon. Not. R. Astron. Soc. {\bf 464}, 3101 (2017); {\bf 479}, 4735 (2018); 
   {\bf 489}, 3022 (2019).

  
\bibitem{AK1996}
   N. Andersson and K. D. Kokkotas, Phys.\ Rev.\ Lett.\ {\bf 77}, 4134 (1996).

\bibitem{AK1998}
   N. Andersson and K. D. Kokkotas, Mon.\ Not.\ R. Astron.\ Soc.\ {\bf 299}, 1059 (1998).

\bibitem{STM2001}
   H. Sotani, K. Tominaga, and K. I. Maeda, Phys.\ Rev.\ D {\bf 65}, 024010 (2001).

\bibitem{SH2003}
   H. Sotani and T. Harada, Phys.\ Rev.\ D {\bf 68}, 024019 (2003);
   H. Sotani, K. Kohri, and T. Harada, {\it ibid}.\ {\bf 69}, 084008 (2004).

\bibitem{TL2005}
   L. K. Tsui and P. T. Leung, Mon.\ Not.\ R. Astron.\ Soc.\ {\bf 357}, 1029 (2005).

\bibitem{SYMT2011}
   H. Sotani, N. Yasutake, T. Maruyama, and T. Tatsumi, Phys.\ Rev.\ D {\bf 83} 024014 (2011).

\bibitem{PA2012}
   A. Passamonti and N. Andersson, Mon.\ Not.\ R. Astron.\ Soc.\ {\bf 419}, 638 (2012).

\bibitem{DGKK2013}
   D. D. Doneva, E. Gaertig, K. D. Kokkotas, and C. Kr\"{u}ger, Phys.\ Rev.\ D {\bf 88}, 044052 (2013).

\bibitem{Sotani20b}
   H. Sotani, Phys. Rev. D {\bf 102}, 063023 (2020); 103021 (2020).

\bibitem{Sotani21}
   H. Sotani, Phys. Rev. D {\bf 103}, 123015 (2021).

\bibitem{FMP2003}
   V. Ferrari, G. Miniutti, and J. A. Pons, Mon. Not. R. Astron. Soc. {\bf 342}, 629 (2003).

\bibitem{FKAO2015}
   J. Fuller, H. Klion, E. Abdikamalov, and C. D. Ott, Mon.\ Not.\ R. Astron.\ Soc.\ {\bf 450}, 414 (2015).

\bibitem{ST2016}
   H. Sotani and T. Takiwaki, Phys.\ Rev.\ D {\bf 94}, 044043 (2016); {\bf 102}, 023028 (2020); 
   Mon. Not. R. Astron. Soc. {\bf 498}, 3503 (2020).

\bibitem{SKTK2017}
   H. Sotani, T. Kuroda, T. Takiwaki, and K. Kotake, Phys.\ Rev.\ D {\bf 96}, 063005 (2017).

\bibitem{MRBV2018}
  V. Morozova, D. Radice, A. Burrows, and D. Vartanyan, Astrophys. J. {\bf 861}, 10 (2018).

\bibitem{SKTK2019}
   H. Sotani, T. Kuroda, T. Takiwaki, and K. Kotake, Phys.\ Rev.\ D {\bf 99}, 123024 (2019).

\bibitem{TCPOF19}
   A. Torres-Forn\'{e}, P. Cerd\'{a}-Dur\'{a}n, A. Passamonti, M. Obergaulinger, and J. A. Font, Mon. Not. R. Astron. Soc. {\bf 482}, 3967 (2019).

\bibitem{SS2019}
   H. Sotani and K. Sumiyoshi, Phys.\ Rev.\ D {\bf 100}, 083008 (2019); Mon. Not. R. Astron. Soc. {\bf 507}, 2766 (2021).

\bibitem{ST2020}
   H. Sotani and T. Takiwaki, Phys.\ Rev.\ D {\bf 102}, 063025 (2020).

\bibitem{STT2021}
   H. Sotani, T. Takiwaki, and H. Togashi, preprint (arXiv:2110.03131).

\bibitem{KHA15}
   C. J. Kr\"{u}ger, W. C. G. Ho, and N. Andersson, Phys.\ Rev.\ D {\bf 92}, 063009 (2015).

\bibitem{Page2020}
D.~{Page}, M.~V.~{Beznogov}, I.~{Garibay}, J.~M.~{Lattimer}, M.~{Prakash}, and H.~T.~{Janka}, Astrophys. J. {\bf 898}, 125 (2020). 
\bibitem{Lattimer1994}
J.~M.~{Lattimer}, K.~A.~{Riper}, M.~{Prakash}, M.~{Prakash}, Astrophys. J. {\bf 425}, 802 (1994). 
\bibitem{Gnedin2001}
O.~Y.~{Gnedin}, D.~G.~{Yakovlev}, and A.~Y.~{Potekhin}, Mon. Not. R. Astron. Soc. {\bf 324}, 725 (2001).
\bibitem{Sales2020}
T.~{Sales}, O.~{Louren{\c{c}}o}, M.~{Dutra}, and R.~{Negreiros}, Astron. Astrophys {\bf 642}, 42 (2020).
\bibitem{YP04}  
   D. G. Yakovlev and C. J. Pethick, Annu. Rev. Astron. Astrophys. J. {\bf 42}, 169 (2004)
\bibitem{Page06} 
   D. Page, U. Geppert, and F. Weber, Nucl. Phys. A {\bf 777}, 497 (2006).
\bibitem{Burgio2021}
G.~F.~Burgio, H.~J.~Schulze, I.~Vida\~{n}a, and J.~B.~Wei, Prog. Part. Nucl. Phys. {\bf 120}, 103879 (2021).

\bibitem{Page2011}
   D. Page, M. Prakash, J. M. Lattimer, and A. W. Steiner, Phys. Rev. Lett. {\bf 106}, 081101 (2011).

\bibitem{Shternin2011}
   P. S. Shternin, D. G. Yakovlev, C. O. Heinke, W. C. G. Ho, D. J. Patnaude, Mon. Not. Roy. Astron. Soc. {\bf 412}, L108  (2011).
   
   \bibitem{Page2004}
   D. Page, J. M. Lattimer,  M. Prakash and A. W. Steiner, Astrophys. J. Suppl. {\bf 155}, 623 (2004).



   

   
\bibitem{TM1e}  
   H. Shen, F. Ji, J. Hu, and K. Sumiyoshi, Astrophys. J. {\bf 891}, 148 (2020).

\bibitem{Togashi17}  
   H. Togashi, K. Nakazato, Y. Takehara, S. Yamamuro, H. Suzuki, and M. Takano, Nucl. Phys. A {\bf 961}, 78 (2017).

\bibitem{Shen}  
   H. Shen, H. Toki, K. Oyamatsu, and K. Sumiyoshi, Nucl. Phys. {\bf A637}, 435 (1998).

   
\bibitem{SIOO14} 
   H. Sotani, K. Iida, K. Oyamatsu, and A. Ohnishi, Prog. Theor. Exp. Phys. {\bf 2014}, 051E01 (2014).
  

\bibitem{SSB16} 
   H. O. Silva, H. Sotani, and E. Berti, Mon. Not. R. Astron. Soc. {\bf 459}, 4378 (2016).   

\bibitem{Sotani17}  
   H. Sotani, Phys. Rev. C {\bf 95}, 025802 (2017).

\bibitem{SK17}  
   H. Sotani and K. D. Kokkotas, Phys. Rev. D {\bf 95}, 044032 (2017).
   




\bibitem{Blaschke20}  
   D. Blaschke, A. Ayriyan, D. E. Alvarez-Castillo, and H. Grigorian, universe {\bf 6}, 81 (2020).

\bibitem{Fujimoto84}  
   M. Y. Fujimoto, T. Hanawa, I. Iben, Jr., and M. B.  Richardson, Astrophys. J.  {\bf 278},  813 (1984)

\bibitem{Dohi19}  
   A. Dohi, K. Nakazato, M. Hashimoto, Y. Matsuo, and T. Noda, Prog. Theor. Exp. Phys. {\bf 2019}, 113E01 (2019).

\bibitem{Dohi21} 
   A. Dohi, H.-L. Liu, T. Noda, M. Hashimoto, arXiv: 2112.13302.
   
\bibitem{Schatz1999}
H.~{Schatz}, L.~{Bildsten}, A.~{Cumming},  and M.~{Wiescher}, Astrophys. J. {\bf 524},
  1014 (1999).
\bibitem{Potekhin2015}
A.Y. {Potekhin}, J.A. {Pons},  and D.~{Page}, Space Sci. Rev. {\bf 191}, 239 (2015).
\bibitem{Baiko2001}
D.A. {Baiko}, P.~{Haensel},  and D.G. {Yakovlev}, Astron. Astrophys. J. {\bf 374}, 151 (2001).

\bibitem{Prakash92} M.~Prakash, M.~Prakash, J.~M.~Lattimer, C.~J.~Pethick, Astrophys. J. {\bf 390}, L77 (1992).
\bibitem{Tsuruta09} S. Tsuruta, J. Sadino, A. Kobelski, M. A. Teter, A. C. Liebmann, T. Takatsuka, K. Nomoto, and H. Umeda, Astrophys. J. {\bf 691}, 621 (2009).
\bibitem{Raduta18} A.~R.~Raduta , A.~Sedrakian, F.~Weber, Mon. Not. R. Astron. Soc. {\bf 475}, 4347 (2018).

\bibitem{BPR20}  
   M. V. Beznogov, D. Page, and E. Ramirez-Ruiz, Astrophys. J.  {\bf 888},  97 (2020).

\bibitem{LHL17}  
   Y. Lim, C. H. Hyun, and C. -H. Lee, Int. J. Mod. Phys. E 26, 1750015 (2017).

\bibitem{LPPH91}  
   J. M. Lattimer, C. J. Pethick, M. Prakash, and P. Haensel, Phys. Rev. Lett. 66, 2701 (1991).
   
\bibitem{Gusakov05}
   M. E. Gusakov, A. D. Kaminker, D. G. Yakovlev, and O. Y. Gnedin, Mon. Not. R. Astron. Soc. {\bf 363}, 555 (2005).
   
\bibitem{Shternin11}
   P. S. Shternin, D. G. Yakovlev, C. O. Heinke, W. C. G. Ho, and D. J. Patnaude, Mon. Not. R. Astron. Soc. {\bf 412}, L108 (2011).
   
\bibitem{Gudmundsson1983}
E.~H.~Gudmundsson, C.~J.~Pethick, and R.~I.~Epstein, Astrophys. J.  {\bf 272}, 286 (1983).
\end{thebibliography}

\end{document}